\documentclass{PoS}
\usepackage{amsmath}
\usepackage{slashed}
\newcommand\beq{\begin{eqnarray}}
\newcommand\eeq{\end{eqnarray}}
\newcommand\nn{\nonumber}
\newcommand\Eq[1]{Eq.~(\ref{eq:#1})}
\newcommand\Sec[1]{Sec.~\ref{sec:#1}}

\title{Background field method and nonrelativistic QED matching}

\ShortTitle{Background field method and NRQED matching}

\author{\speaker{Jong-Wan Lee}
\thanks{Work supported in part by a joint City College of New York-RIKEN/BNL Research Center fellowship, 
an award from the Professional Staff Congress of the City University of New York, 
and by the U.S. National Science Foundation, 
under Grant No. PHY12-05778.}\\
        Department of Physics, The City College of New York, New York, NY 10031, USA\\
        E-mail: \email{jwlee2@ccny.cuny.edu}}

\author{Brian C. Tiburzi\\
        Department of Physics, The City College of New York, New York, NY 10031, USA\\
        Graduate School and University Center, The City University of New York, New York, NY 10016, USA\\
	RIKEN BNL Research Center, Brookhaven National Laboratory, Upton, NY 11973, USA\\
	E-mail: \email{btiburzi@ccny.cuny.edu}}

\abstract{We discuss the resolution of an inconsistency between lattice background field methods 
and nonrelativistic QED matching conditions. 
In particular, we show that 
lack of on-shell conditions in lattice QCD with time-dependent background fields generally requires 
that certain operators related by equations of motion should be retained 
in an effective field theory to correctly describe the behavior of Green's functions. 
The coefficients of such operators in a nonrelativistic hadronic theory 
are determined by performing a robust nonrelativistic expansion of 
QED for relativistic scalar and spin-half hadrons including nonminimal electromagnetic couplings. 
Provided that nonrelativistic QED is augmented with equation-of-motion operators, 
we find that the background field method can be reconciled with the nonrelativistic QED matching 
conditions without any inconsistency. 
We further investigate whether nonrelativistic QED can be employed 
in the analysis of lattice QCD correlation function in background fields, 
but we are confronted with difficulties. 
Instead, we argue that the most desirable approach is a hybrid one 
which relies on a relativistic hadronic theory with operators chosen 
based on their relevance in the nonrelativistic limit. 
Using this hybrid framework, we obtain practically useful forms of 
correlation functions for scalar and spin-half hadrons in uniform electric and magnetic fields. 
}

\FullConference{The 32nd International Symposium on Lattice Field Theory,\\
		23-28 June, 2014\\
		Columbia University New York, NY}

\begin{document}
\section{Introduction}
The lattice background field method provides a first principle approach 
to the calculation of 
low-energy hadronic properties such as magnetic moments and electromagnetic 
polarizabilities \cite{Martinelli1982}. 
The power of this method becomes striking when it is incorporated with 
an effective field theory, namely nonrelativistic QED (NRQED); 
the universal low-energy parameters in NRQED, 
which encompass hadronic structure determined from scattering experiments, 
can be computed using lattice QCD in electromagnetic fields. 
Although the standard NRQED matching of $S$-matrix elements relates 
these low-energy parameters to physical observables without any problem, 
however, the extension of matching to theories in background fields 
becomes subtle.  

To point out the physical relevance of this subtlety, 
we consider the energy shift of a charged spin-half hadron in the zero momentum limit 
for the case of a uniform electric field, 
\beq
\Delta E(\vec{E})=
\left(c_{A_1}+\frac{1}{2}c_{A_2}\right)\frac{\vec{E}^2}{8M^3}=
-\frac{1}{2}\left[4\pi \alpha_E-\frac{Z^2+\kappa^2}{4M^3}
-\frac{Z}{3M}\langle r_E^2\rangle\right]\vec{E}^2,
\label{eq:energy_shift}
\eeq
where $M$, $\kappa$, $\alpha_E$, and $\langle r_E^2\rangle$ are the hadron mass, 
anomalous magnetic moment, 
electric polarizability, and charge radius, respectively. 
The parameters $c_{A_1}$ and $c_{A_2}$ are the coefficients 
of terms quadratic in electric and magnetic fields 
in the NRQED Lagrangian, \Eq{nrqed}, and 
have been determined from the standard NRQED matching \cite{Hill2013}. 
While the real Compton scattering process in NRQED is independent from $\langle r_E^2\rangle$, 
the NRQED action for proton in a uniform electric field shows an energy shift as in \Eq{energy_shift}. 
The appearance of the charge radius, which can only arise from virtual processes, 
is surprising.\footnote{
For a charged scalar hadron, we find a similar situation, 
where the initial energy shift in a uniform electric field 
receives a contribution from the charge radius \cite{Lee2014_1}:
$
\Delta E(\vec{E})=
\left(c_{A_1}+\frac{1}{2}c_{A_2}\right)\frac{\vec{E}^2}{8M^3}=
-\frac{1}{2}\left[4\pi \alpha_E+\frac{Z^2}{2M^3}
-\frac{Z}{3M}\langle r_E^2\rangle\right]\vec{E}^2.
$
}
We claim this is an inconsistency that can be resolved by 
proper consideration of operators related by equations of motion.
In particular, we demonstrate how the equation-of-motion (EOM) operators 
modify the time dependence of Green's functions 
using a toy model of relativistic effective theory for a scalar hadron 
in time-dependent electromagnetic fields. 
In addition, we discuss the difficulties involved in employing NRQED for 
the analysis of background field correlation functions obtained from lattice QCD, 
and argue that the most practical approach is a hybrid scheme, 
which combines a relativistic hadronic theory with operator selection 
based on their relevance in the nonrelativistic limit. 
For full details, see \cite{Lee2014_1,Lee2014_2}. 

\section{Equation-of-motion operators and (non)relativistic QED}
In formulating an effective field theory, 
the equations of motion 
are exclusively used to economically reduce the number of low-energy parameters by 
eliminating redundant operators. 
The physical consequences of the theory are not lost, 
because a theory containing EOM operators 
and the reduced theory without such operators 
are equivalent, provided that the parameters in each theory 
are determined by matching $S$-matrix elements. 
In external fields, however, 
the equivalence between the full and reduced theory 
becomes subtle at the level of Green's functions, 
where on-shell conditions can be lost. 

To expose this subtlety, consider the toy-model Lagrangian 
for a charged composite scalar
\beq
\mathcal{L}^{\textrm{toy}}_{\textrm{full}}=D_\mu \Phi^\dagger D^\mu \Phi-M^2 \Phi^\dagger \Phi
+\frac{C_1}{2M^4}\Phi^\dagger\Phi \partial^2 F^2
+\frac{C_2}{M^4}F^2(D_\mu\Phi^\dagger D^\mu\Phi-M^2\Phi^\dagger\Phi),
\label{eq:toy_Lagrange}
\eeq
where the gauge covariant derivative is $D_\mu=\partial_\mu-iZA_\mu$, 
and the electromagnetic field strength is given by $F^{\mu\nu}=\partial^\mu A^\nu-\partial^\nu A^\mu$. 
In Minkowski space-time, we adopt the metric $\eta^{\mu\nu}=\delta^{\mu\nu}\{-1,1,1,1\}$. 
No power counting has been utilized in \Eq{toy_Lagrange}, instead 
we merely select the last two operators 
to illustrate our point. 
We also assume that 
the coefficients of these operators are proportional to 
the electromagnetic coupling, $\alpha=\frac{e^2}{4\pi}\ll 1$, 
and we will drop terms of $\mathcal{O}(\alpha^2)$ in what follows. 

We first consider on-shell processes, where 
one can show that observables, such as the Compton amplitude, 
depend only on a particular linear combination of low-energy parameters, $C_1+C_2$. 
For off-shell processes appearing in loop diagrams, 
additional dependence on the $C_2$ can arise; 
however, this must be canceled by the counterterms necessary to renormalize the theory. 
Because the diagramatic approach is cumbersome, 
we instead employ field redefinitions to remove redundant operators, see \cite{Arzt1995} 
and references therein. 
By invoking the field redefinition, 
$\Phi=\left(1-\frac{C_2}{2M^4}F^2\right)\Phi'$, 
which corresponds to dressing the scalar field with photons, 
we obtain the reduced theory
\beq
\mathcal{L}^{\textrm{toy}}_{\textrm{reduced}}=D_\mu \Phi'^\dagger D^\mu \Phi'-M^2 \Phi'^\dagger \Phi'
+\frac{C'_1}{2M^4}\Phi'^\dagger\Phi' \partial^2 F^2+\mathcal{O}(\alpha^2),
\label{eq:toy_reduced}
\eeq
with $C'_1=C_1+C_2$. The coefficient $C'_1$ can be choosen so that \Eq{toy_reduced} 
reproduces $S$-matrix elements for processes involving the composite scalar and photons. 
In this way, the theories described by Eqs.~(\ref{eq:toy_Lagrange}) and (\ref{eq:toy_reduced}) 
are equivalent. 

Now consider that $F^{\mu\nu}$ in \Eq{toy_Lagrange} is a time-dependent external field. 
Since the explicit time dependence prevents the possibility of an on-shell condition 
leading to the absence of single-particle poles, 
one cannot rely on a renormalization prescription to fix the behavior of the two-point function 
at the pole. 
Instead we appeal to the Green's function to resolve the parameters $C_1$ and $C_2$. 
Consider the propagators for $\Phi'$ and $\Phi$ given by 
$\mathcal{G}'(x,y)=\langle 0|T\{\Phi'(x)\Phi'^\dagger(y)\}|0\rangle$ 
and 
$\mathcal{G}(x,y)=\langle 0|T\{\Phi(x)\Phi^\dagger(y)\}|0\rangle$, respectively. 
These propagators are related to each other by 
\beq
\mathcal{G}(x,y)=\left[1-\frac{C_2}{2M^4}\left[F^2(x)+F^2(y)\right]\right] \mathcal{G}'(x,y),
\label{eq:prop}
\eeq
where contributions that are of order $\alpha^2$ are dropped. 
The propagators $\mathcal{G}(x,y)$ and $\mathcal{G}'(x,y)$ generally have 
different time dependence. 
We cannot simply assume that the external field propagator 
will be the one obtained by using the reduced theory, $\mathcal{G}'(x,y)$. 
Because the most general effective theory should have all possible operators, 
we should rather retain operators related by equations of motion. 
Furthermore, we can access to both parameters $C'_1$ and $C_2$ 
from the time dependence of the propagator, \Eq{prop}, 
where the parameter $C'_1$ contributes to on-shell properties of $\Phi$. 
Therefore, we must retain operators ordinarily removed by equations of motion 
when we consider time-dependent external field correlation functions. 

Moving on from the toy-model example, we consider augmented NRQED 
for spin-half hadrons by retaining certain EOM operators. 
We find the Lagrangian up to $\mathcal{O}(M^{-3})$
\beq
\mathcal{L}^{\textrm{NRQED}}&=&\psi^\dagger
\left[iD_0+c_2\frac{\vec{D}^2}{2M}+c_4\frac{\vec{D}^4}{8M^3}
+c_F\frac{\vec{\sigma}\cdot\vec{B}}{2M}+c_D\frac{[\vec{\nabla}\cdot\vec{E}]}{8M^2}
+ic_S\frac{\vec{\sigma}\cdot(\vec{D}\times\vec{E}-\vec{E}\times\vec{D})}{8M^2}
\right. \nn \\
&&
\left.
+c_{W_1}\frac{\{\vec{D}^2,\vec{\sigma}\cdot\vec{B}\}}{8M^3}
-c_{W_2}\frac{D^i \vec{\sigma}\cdot\vec{B} D^i}{4M^3}
+c_{p'p}\frac{ \big\{ \vec{D}\cdot\vec{B}, \vec{\sigma}\cdot\vec{D} \big\}
}{8M^3}
+c_{A_1}\frac{\vec{B}^2-\vec{E}^2}{8M^3}
-c_{A_2}\frac{\vec{E}^2}{16M^3}
\right. \nn \\
&&
\left.
+ic_M\frac{\{D^i,[\vec{\nabla}\times\vec{B}]^i\}}{8M^3}
+c_{X_0}\frac{[iD_0,\vec{D}\cdot\vec{E}+\vec{E}\cdot\vec{D}]}{8M^3}
+c_{X'_0}\frac{[D_0,[D_0,\vec{\sigma}\cdot \vec{B}]]}{8M^3}
\right]\psi,
\label{eq:nrqed}
\eeq
where $D_0=\partial_0-iZA_0$ and $D^i=\nabla^i-iZA^i$. 
Here we impose Hermiticity and invariance under $P$, $T$, and gauge transformations. 
The electric and magnetic fields $\vec{E}$ and $\vec{B}$ are given by 
standard expressions, $\vec{E}=-\partial_0-\vec{\nabla}A^0$ and 
$\vec{B}=\vec{\nabla}\times \vec{A}$, respectively. 
Note that we have adopted the convention that bracketed derivatives 
only act inside the square brackets. 

The last two operators in \Eq{nrqed} are ordinarily redundant, 
and for on-shell processes those operators can be removed using the equations of motion
\beq
\psi^\dagger \frac{[iD_0,\vec{D}\cdot\vec{E}+\vec{E}\cdot\vec{D}]}{8M^3}\psi
\overset{\textrm{eom}}{=}
-c_2\psi^\dagger \frac{[\vec{D}^2,\vec{D}\cdot\vec{E}+\vec{E}\cdot\vec{D}]}{16M^4}\psi. 
\label{eq:eom}
\eeq
Similar results can be shown for the operator having coefficient $c_{X'_0}$, 
and what remains is the $\mathcal{O}(M^{-3})$ standard NRQED Lagrangian. 
Off shell, however, these operators can modify Green's functions 
and need to be accounted for to describe lattice QCD correlations functions 
in external fields. For example, for uniform electric fields 
the equality in \Eq{eom} is no longer valid; 
the right-hand side is zero, while 
the left-hand side is propotional to $Z\vec{E}^2$. 
The non-zero $c_{X_0}$ term modifies the time dependence of the Green's function obtained 
by using \Eq{nrqed} in an essential way. 
We stress that no modification is necessary to standard NRQED matching conditions 
that utilize scattering amplitudes. 

The coefficients $c_{X_0}$ and $c_{X'_0}$ cannot be determined from 
the standard one- and two-photon matching combined even with the implementation 
of Lorentz invariance \cite{Hill2013,Heinonen2012}. Instead, we find that these coefficients 
can be resolved at the level of the relativistic effective theory, 
where EOM operators 
turn out to be innocuous. 
For a relativistic charged spin-half hadron, we have \cite{Lee2014_2}
\beq
\mathcal{L}^{\textrm{spin-half}}&=&\bar{\Psi}\left[
i\slashed{D}-M+\frac{\kappa}{4M}\sigma_{\mu\nu}F^{\mu\nu}
-\frac{\mathcal{C}_1}{M^2}\gamma_\mu[D_\nu,F^{\mu\nu}]
+\frac{\mathcal{C}_2}{M^3}\sigma_{\mu\nu}[D_\rho,[D^\mu,F^{\nu\rho}]]
+\frac{\mathcal{C}_3}{M^3}F^{\mu\nu}F_{\mu\nu}\right]\Psi
\nn \\
&&+\frac{i\mathcal{C}_4}{M^4}\bar{\Psi}\gamma_\mu D_\nu \Psi T^{\mu\nu}
-\frac{\mathcal{C}_5}{M^5}D_\mu \bar{\Psi} D_\nu \Psi T^{\mu\nu}
-\frac{\mathcal{C}_6}{M^5}D_\rho \bar{\Psi} D^\rho \Psi F^{\mu\nu}F_{\mu\nu},
\label{eq:spin_half}
\eeq
where the matrix $\sigma_{\mu\nu}$ is defined as $\sigma_{\mu\nu}=\frac{i}{2}[\gamma_\mu,\gamma_\nu]$ 
as usual. Notice that the last three operators will be relevant to $\mathcal{O}(M^{-3})$ 
in the relativistic limit, because they contain time derivatives acting on the massive hadron field. 
After the phase transformation, $\Psi\rightarrow e^{-iMx_0}\Psi$, 
the nonrelativistic QED action in \Eq{nrqed} is obtained by performing 
a series of Foldy-Wouthuysen (FW) transformations \cite{Foldy1950}. 
Combined with the one- and two-photon matching 
in the relativistic theory, we find 
\beq
c_{X_0}=-\frac{1}{2}\left(c_D-Z-2\kappa\right)=\kappa-\frac{2}{3}M^2 \langle r_E^2\rangle,
~\textrm{and},~
c_{X'_0}=\frac{\kappa}{2}-c_{W_2}=-\frac{2}{3}M^2\langle r_M^2\rangle,
\eeq
where $\langle r_M^2 \rangle$ is the magnetic radius. 
Our results for other parameters agree with the standard NRQED matching conditions \cite{Hill2013}. 
By taking these results into account, 
we obtain 
\beq
\Delta E(\vec{E})=-\frac{1}{2}\left[4\pi\alpha_E-\frac{(Z+\kappa)^2}{4M^3}\right]\vec{E}^2. 
\eeq
We immediately realize that the energy shift is free from the charge radius, 
$\langle r_E^2\rangle$, and thus conclude that there is no inconsistency 
between the background field method and the NRQED matching 
provided EOM operators are retained.\footnote{
Similarly, we can show that there is no inconsistency for the case of 
a scalar hadron. 
Consider the Lagrangian density for a relativistic charged composite scalar \cite{Lee2014_1}
\beq
\mathcal{L}^{\textrm{scalar}}=D_\mu \Phi^\dagger D^\mu\Phi
-M^2\Phi^\dagger\Phi-\frac{C_0}{M^2}F^{\mu\nu}F_{\mu\nu}\Phi^\dagger\Phi
+\frac{C_1}{M^2}[\partial_\mu F^{\mu\nu}]J_\nu
+\frac{C_2}{M^4} T_{\mu\nu} D^\mu \Phi^\dagger D^{\nu} \Phi
-\frac{C_3}{M^4}[\partial^2\partial_\mu F^{\mu\nu}]J_\nu,
\nn
\eeq
where the electromagnetic stress-energy tensor is $T_{\mu\nu}=F_{\rho\{\mu} F_{\nu\}}^{~\rho}$ 
and the vector current is $J_\mu=i(\Phi^\dagger [D_\mu\Phi]-[D_\mu \Phi^\dagger]\Phi)$. 
Using the relation between the relativistic scalar field $\Phi$ 
and nonrelativistic scalar field $\phi$, $\Phi(x)=\frac{e^{-iMt}}{[4(M^2-\vec{D}^2)]^{1/4}}\phi(x)$, 
we perform the $1/M$ expansion of the above Lagrangian by keeping all terms up to $\mathcal{O}(M^{-3})$ 
and arrive at the scalar version of \Eq{nrqed} with
\beq
c_{X_0}=-\frac{1}{2}c_D-Z=-\frac{2}{3}M^2\langle r_E^2\rangle-Z.
\nn
\eeq
In this derivation, field redefinitions are carefully carried out to 
preserve the time dependence of Green's functions. 
The same result can also be obtained by matching the 
relativistic Green's functions to the nonrelativistic ones. 
In uniform electric fields, the $c_{X_0}$ term makes a nonvanishing contribution 
to the energy shift and we find
\beq
\Delta E(\vec{E})=-\frac{1}{2}\left[
4\pi\alpha_E+\frac{Z^2}{2M^3}
\right]\vec{E}^2,
\nn
\eeq
where the contribution from the charge radius, $\langle r_E^2\rangle$, 
is absent due to inclusion of the EOM operator.
}

\section{Euclidean correlation functions}
Nonrelativistic QED provides an ideal framework to compute single-hadron 
propagators at low energy because of the relatively simple form of interactions 
and the manifest power counting. 
However, the correlation functions determined with lattice QCD 
are necessarily relativistic, and in general the comparison with NRQED predictions
is highly nontrivial; 
one must transform the lattice correlators to nonrelativistic ones, 
i.e. the FW transformation for spin-half hadrons, and the normalization factor 
for scalar hadrons which has been used to relate the relativistic field $\Phi$ 
to the nonrelativistic one $\phi$. 
To perform this transformation, one requires knowledge of hadronic parameters and must 
accordingly propagate uncertainties through the transformation. 

A more desirable approach is to compute correlation functions using 
relativistic effective theory with operators chosen by their relevance 
in the nonrelativistic limit, 
where one avoids the lack of manifest power counting of the relativistic theory. 
Of course, this relativistic approach is consistent with the NRQED 
in terms of matching. 
Indeed, this semirelativistic philosophy is employed to 
write down the relativistic theory for a charged spin-half hadron in \Eq{spin_half}. 
For uniform external fields, the action is further reduced to 
\footnote{Throughout this section all calculations are performed in Euclidean space-time 
with the metric $\eta_E^{\mu\nu}=\delta^{\mu\nu}$. 
For instance, the Euclidean electric field $\vec{\mathcal{E}}$ 
is related to the Minkowski electric field $\vec{E}$ by 
the analytic continuation $\vec{\mathcal{E}}=-i\vec{E}$.}
\beq
\mathcal{L}^{\textrm{spin-half}}_{\textrm{reduced}}=\bar{\Psi}\left[
\slashed{D}+\mathcal{M}_{\mathcal{E} (B)}+\frac{\kappa}{2M}\sigma_{\mu\nu} F_{\mu\nu}
\right]\Psi,
\eeq
where the mass parameters 
are given as 
$\mathcal{M}_\mathcal{E}=M+\frac{1}{2}4\pi\alpha_E\mathcal{E}^2$ and 
$\mathcal{M}_B=M-\frac{1}{2}4\pi\beta_M B^2$. 
We neglect higher-order effects from the electromagnetic field in 
$\mathcal{M}_\mathcal{E}$, $\mathcal{M}_B$, and $\kappa$. 

\subsection{Uniform electric field}
\label{sec:uniform_E}
A charged hadron in a uniform electric field 
does not have definite energy eigenstates 
and the correlation function will exhibit nonstandard time dependence. 
The integral form of a boost-projected correlation function can be found 
in Ref. \cite{Detmold2010}, which results in cumbersome numerical fits to lattice QCD data. 
To avoid this complication, we consider the nonrelativistic limit of 
the correlation function using the velocity power counting of NRQED. 
The $\mathcal{O}(v^4)$ charged scalar propagator is given by
\beq
G^{\mathcal{E}}(\tau)=
e^{-\eta-\frac{\zeta^2}{6\eta}}\left[
1-\frac{\zeta^2}{4\eta^2}\left(1-\frac{\zeta^2}{10\eta}\right)
-\frac{\zeta^2}{4\eta^3}\left(1-\frac{5\zeta^2}{8\eta}+\frac{17\zeta^4}{280\eta^2}
-\frac{\zeta^6}{800\eta^3}\right)
\right],
\label{eq:nr_corr_E}
\eeq
where $\eta=\mathcal{M}_{\mathcal{E}}\tau\sim\mathcal{O}(v^{-2})$ 
and $\zeta=Z\mathcal{E}\tau^2\sim\mathcal{O}(v^{-1})$. For a charged spin-half hadron, the 
boosted-projected correlation functions take the form, 
\beq
G^{\mathcal{E}}_\pm(\tau)=\left(1\pm\frac{\kappa\mathcal{E}}{2M^2}\right)
G^{\mathcal{E}}(\tau),
\label{eq:corr_E}
\eeq
with $\eta$ replaced by $\eta=\sqrt{\mathcal{M}_\mathcal{E}^2
-\frac{\kappa^2\mathcal{E}^2}{4M^2}\pm Z\mathcal{E}}\,\tau$. 
The neutral scalar and spin-half propagators can be obtained from 
Eqs.~(\ref{eq:nr_corr_E}) and (\ref{eq:corr_E}) 
by taking vanishing electric charge ($Z=\zeta=0$), respectively. 

\subsection{Uniform magnetic field}
\label{sec:uniform_B}
In contrast to the uniform electric field, there is no explicit time dependence 
for a uniform magnetic field, which implies the existence of definite energy eigenstates. 
However, another complication arises from the fact that 
a space-averaged correlation function receives contributions from 
an infinite tower of Landau levels; 
the characteristic energy splitting between adjacent Landau levels is 
$\Delta E=|ZB|/M$ for weak external fields, and thus 
the correlation function will suffer from significant excited-state contamination. 
Therefore, it becomes important to isolate the ground state by projecting the 
correlator onto the lowest Landau level using a ground-state harmonic 
oscillator wave function, 
where in principle we can generalize this projection to an arbitrary $n$th Landau level 
\cite{Tiburzi2013}. 
For a charged scalar, the propagator projected onto the $n$th Landau level has the simple form, 
$G^B_n(\tau)=Z_n e^{-E_n \tau}$, assuming that $\tau>0$. The energy $E_n$ is given by
\beq
E_n=\sqrt{M^2+|ZB|(2n+1)-4\pi\beta_M MB^2}.
\label{eq:scalar_energy_B}
\eeq
For a charged spin-half hadron 
we further perform spin and parity projections denoted by $\pm_1$ and $\pm_2$, respectively, 
and obtain the propagator, 
$G^B_{(n,\pm_1,\pm_2)}(\tau)=Z_{(n,\pm_1,\pm_2)}e^{-E_{(n,\pm_1,\pm_2)}\tau}$, 
using Schwinger's proper-time trick \cite{Schwinger1951,Tiburzi2008}, 
where the amplitudes and energies of these four eigenstates are given by
\beq
Z_{(n,\pm_1,\pm_2)}&=&\frac{1}{2}\left[
\frac{\mathcal{M}_B}{\sqrt{\mathcal{M}_B^2\mp_1 ZB + |ZB|(2n+1)}}
\pm_2 1
\right],
\\
E_{(n,\pm_1,\pm_2)}&=&\sqrt{\mathcal{M}_B^2\mp_1 ZB + |ZB|(2n+1)}\mp_1
\left(\pm_2\frac{\kappa B}{2M}\right).
\label{eq:spinor_energy_B}
\eeq
For a given Landau level, there are four distinct positive-energy eigenstates 
disentangled by using spin and parity projectors, 
and each projected correlator decays with a simple exponential in Euclidean time. 
Notice that the negative-parity amplitudes, $Z_{(n,\pm,-)}$, scale as $ZB/M^2$ 
and vanish in the nonrelativistic limit, as promised. 
From expanding energies in the weak magnetic field limit, 
we find the first relativistic correction to the Zeeman splitting, 
$E_\uparrow-E_\downarrow=-\frac{B}{M} \left[\mu-Z\frac{|ZB|}{M^2}\left(n+\frac{1}{2}\right)\right]$, 
and previously overlooked $B^2$ terms which contribute to the spin-averaged energy, 
$\frac{1}{2}(E_\uparrow+E_\downarrow)=M+\frac{|ZB|}{M}\left(n+\frac{1}{2}\right)
\left[1-\frac{|ZB|}{2M^2}\left(n+\frac{1}{2}\right)\right]-\frac{1}{2}
\left(4\pi\beta_M+\frac{Z^2}{4M^3}\right)B^2$.
The results for neutral scalar and spin-half hadrons can be obtained 
from Eqs.~(\ref{eq:scalar_energy_B})--(\ref{eq:spinor_energy_B}) 
by taking vanishing electric charge, respectively. 

\section{Summary}
Above we show that an inconsistency between standard NRQED for hadrons and 
the lattice background field method can be resolved by augmenting NRQED with 
equation-of-motion operators which are normally redundant for on-shell processes. 
We determine the coefficients of such operators by performing a 
brute-force nonrelativistic expansion of the underlying relativistic Lagrange density.
After we discussed difficulties in utilizing NRQED for 
the analysis of lattice QCD data, 
we argue that a desirable approach 
is the hybrid scheme based on 
a relativistic hadron effective theory with operators selected 
by employing NRQED power counting, 
where a direct comparison of lattice correlation functions is possible. 
Using this hybrid approach, we find that the relativistic effective theory for hadrons 
in uniform external fields takes a simple form in which 
the low-energy constants are directly related with physical observables, 
i.e. magnetic moment and electromagnetic polarizabilities.
Therefore, the background field correlation functions 
presented in \Sec{uniform_E} and \Sec{uniform_B} 
will be useful in extracting hadronic parameters from lattice QCD simulations 
with uniform electric and magnetic fields. 

Before we conclude our work, we want to discuss the applicability of our argument to 
the investigation of the spin response of a nucleon, 
where the relevant operators are given as
\beq
\mathcal{L}^{\textrm{spin}}_{eff}=
-\Psi^\dagger2\pi i\left[
-\gamma_{E_1E_1}\vec{\sigma}\cdot\vec{\mathcal{E}}\times\dot{\vec{\mathcal{E}}}
+\gamma_{M_1M_1}\vec{\sigma}\cdot\vec{B}\times\dot{\vec{B}}
+\gamma_{M_1E_2}\sigma^i \mathcal{E}^{ij} B^j
+\gamma_{E_1M_2}\sigma^i B^{ij} \mathcal{E}^j
\right]\Psi,
\eeq
where $\dot{X}=\frac{X}{\partial\tau}$ denotes the Euclidean time derivative 
and $X^{ij}=\frac{1}{2}(\partial^i X^j+\partial^j X^i)$. 
The four constants $\gamma_{E_1E_1}$, $\gamma_{M_1M_1}$, $\gamma_{M_1E_2}$, 
and $\gamma_{E_1M_2}$ are the spin polarizabilities. 
In contrast to the case of uniform electromagnetic fields, 
unfortunately, inclusion of nonuniform electromagnetic fields explicitly 
modifies Green's functions even in a relativistic theory. 
To determine the spin polarizabilities, therefore, 
one must understand which EOM operators are relevant to 
nonuniform electric and magnetic fields, and how 
these operators modify the time dependence of 
background field correlation functions. 



\begin{thebibliography}{99}
\bibitem{Martinelli1982} G. Martinelli, G. Parisi, R. Petronzio, and F. Rapuano, 
Phys. Lett. {\bf 116B}, 434 (1982); C.W. Bernard, T. Draper, K. Olynyk, and M. Rushton, 
Phys. Rev. Lett. {\bf 49}, 1076 (1982); H. Fiebig, W. Wilcox, and R. Woloshyn, Nucl. Phys. {\bf B324}, 
47 (1989). 

\bibitem{Hill2013} R.J. Hill, G. Lee, G. Paz, and M.P. Solon, Phys. Rev. D {\bf 87}, 053017 (2013).

\bibitem{Lee2014_1} J.-W. Lee and B.C. Tiburzi, Phys. Rev. D {\bf 89}, 054017 (2014). 

\bibitem{Lee2014_2} J.-W. Lee and B.C. Tiburzi, Phys. Rev. D {\bf 90}, 074036 (2014). 


\bibitem{Arzt1995} C. Arzt, Phys. Lett. B {\bf 342}, 189 (1995). 

\bibitem{Heinonen2012} J. Heinonen, R.J. Hill, and M.P. Solon, Phys. Rev. D {\bf 86} 094020 (2012). 

\bibitem{Foldy1950} L.L Foldy and S.A. Wouthuysen, Phys. Rev. {\bf 78}, 29 (1950).

\bibitem{Detmold2010} W. Detmold, B. Tiburzi, and A. Walker-Loud, Phys. Rev. D {\bf 81}, 054502 (2010). 

\bibitem{Tiburzi2013} B. Tiburzi and S. Vayl, Phys. Rev. D {\bf 87}, 054507 (2013). 

\bibitem{Schwinger1951} J.S. Schwinger, Phys. Rev. {\bf 82} 664 (1951).

\bibitem{Tiburzi2008} B.C. Tiburzi, Nucl. Phys. {\bf A814}, 74 (2008).

\end{thebibliography}
\end{document}